\documentclass[a4paper,11pt]{article}
\usepackage{pos}
\usepackage{axodraw2}
\def\O{\mathcal{O}}
\def\nn{\nonumber}
\def\L{\mathcal{L}}
\title{Renormalisation of singlet operators to four loops}

\author*[a]{Giulio Falcioni}

\affiliation[a]{Higgs Centre for Theoretical Physics,\\
  School of Physics and Astronomy, The University of Edinburgh, Edinburgh EH9 3FD, Scotland, UK}

\emailAdd{giulio.falcioni@ed.ac.uk}

\abstract{In QCD the anomalous dimensions of gauge invariant operators of twist 2 play a key role, because they control the scale dependence of the parton distribution functions. Notably, the flavour singlet operators, such as those associated to the gluon distribution, mix under renormalisation with a set of unphysical operators, also known as aliens. Missing this effect leads to wrong results already for the two-loop anomalous dimensions.
The correct renormalisation of gluonic operators is an important step towards the computation of the scale evolution of flavour singlet parton distributions, which is now required to 4 loops.
Leveraging both the background field method and an enhanced BRST symmetry, we construct the required ghost and alien operator basis up to 4 loops for arbitrary mass dimensions. Furthermore, we extract anomalous dimensions at 4 loops, for the physical operators of mass-dimension 4 and 6, and at 3 loops for mass-dimension 8.}

\FullConference{%
  Loops and Legs in Quantum Field Theory - LL2022,\\
  25-30 April, 2022\\
  Ettal, Germany
}

\begin{document}
\maketitle

\section{Introduction}
The increasing precision of the experiments at the Large Hadron Collider (LHC) provided a strong motivation for pushing the theoretical predictions to very high accuracy. Recently, there has been impressive progress in computing perturbative corrections through the next-to-next-to-next-to-leading order (N3LO) in QCD, e.g. \cite{Anastasiou:2015vya,Anastasiou:2016cez,Mistlberger:2018etf,Duhr:2019kwi,Duhr:2020seh,Chen:2021isd,Duhr:2020sdp}.
The lack of knowledge of the 4-loop splitting functions, which govern the scale evolution of parton densities, is one of the dominant sources of theoretical uncertainties on N3LO predictions. Out of a wide range of methods that have been applied to compute splitting functions up to three loops \cite{Floratos:1978ny,Gonzalez-Arroyo:1979qht,Furmanski:1980cm,Hamberg:1991qt,Vogelsang:1995vh,Mertig:1995ny,Ellis:1996nn,Matiounine:1998ky,Matiounine:1998re,Larin:1993vu,Larin:1996wd,Moch:2004pa,Vogt:2004mw,Ablinger:2014nga,Ablinger:2017tan,Behring:2019tus,Ablinger:2019etw,Blumlein:2021enk,Blumlein:2021ryt}, only few approaches are suitable to carry on through four loops. The framework based on the Operator Product Expansion (OPE), originally developed in refs. \cite{Gross:1974cs,Georgi:1974wnj}, is one of the most promising. This approach was applied to determine a highly accurate numerical approximation of the flavour non-singlet splitting functions at four loops and to reconstruct their exact analytic form\footnote{The analytic form of the leading and of the sub-leading contributions to the splitting functions for a large number of {\textit{flavours}} was done in \cite{Gracey:1996ad,Davies:2016jie}.} in the limit of large number of colours \cite{Moch:2017uml}. 

The key point of the method is that the splitting functions are extracted in {\textit{Mellin space}}, from the anomalous dimensions of leading-twist operators, defined as
\begin{equation}
\label{eq:1}
    \mu^2\frac{d}{d\mu^2}\,{\mathcal{O}}^{(N)}_{i;\mu_1\dots\mu_N}  = -\gamma_{ij}^{(N)}\, {\mathcal{O}}^{(N)}_{j;\mu_1\dots\mu_N}.
\end{equation}
Here, the anomalous dimensions $\gamma_{ij}^{(N)}$ give the $N$\textsuperscript{th} Mellin moment of the splitting functions when $i,j=$ q(quark) or g(gluon). These are associated to the renormalisation of the gauge invariant operators
\begin{align}
\label{GIgluOP}
    \O^{(N)}_{g;\mu_1\dots\mu_N}&=\frac{1}{2}\,\mathcal{S}_T\left\{F_{\rho\mu_1}^{a_1}\,D_{\mu_2}^{a_1a_2}\,\dots\,D_{\mu_{N-1}}^{a_{N-2}a_{N-1}}\,F^{a_N;\rho}_{\phantom{a_N;\rho}\mu_N}\right\},\\
\label{GIqS}    
    \O^{(N)}_{q\,\mathrm{S};\mu_1\dots\mu_N}&=\mathcal{S}_T\left\{\bar{\psi}_{i_1}\,\gamma_{\mu_1}\,D_{\mu_2}^{i_1i_2}\,\dots\,D_{\mu_N}^{i_{N-1}i_N}\,\psi_{i_N}\right\},\\
\label{GIqNS}    
    \O^{(N),\rho}_{q\,\mathrm{NS};\mu_1\dots\mu_N}&=\mathcal{S}_T\left\{\bar{\psi}_{i_1}\,(\lambda^\rho)\,\gamma_{\mu_1}\,D_{\mu_2}^{i_1i_2}\,\dots\, D_{\mu_N}^{i_{N-1}i_N}\,\psi_{i_N}\right\},
\end{align}
where $F^a_{\mu\nu}$ is the gluon field-strength, $\psi_i$ is the fermion field, $D_\mu^{ab}$ is the covariant derivative (with colour indices either in the fundamental representation, when it acts on the quark field, or in the adjoint representation when it acts on the field strength) and $\lambda^\rho$ is the generator of the $\mathrm{SU}(n_f)$ flavour group. The symbol $\mathcal{S}_T$ denotes symmetrisation of the Lorentz indices $\mu_1\dots\mu_N$ and removal of trace terms. The determination of the anomalous dimensions is relatively straight forward in the case of the {\textit{flavor non-singlet}} operator, defined in eq.~(\ref{GIqNS}), which renormalise with a multiplicative constant 
\begin{equation}
\O^{(N)}_{q\,\mathrm{NS};\mu_1\dots\mu_N}(\mu^2) = Z_{\mathrm{ns}}^{(N)}(\mu^2)\,\O^{(N),\text{bare}}_{q\,\mathrm{NS};\mu_1\dots\mu_N},\quad\gamma^{(N)}_{\mathrm{ns}} = -\mu^2\frac{d}{d\mu^2}\log Z^{(N)}_{\mathrm{ns}}.
\end{equation}
As a consequence, the non-singlet anomalous dimensions are extracted from {\textit{off-shell}} Operator Matrix Elements (OMEs) with two external fields, which are computed through four loops with state-of-the-art techniques, such as \verb+FORCER+ \cite{Ruijl:2017cxj}. 

The renormalisation of flavour singlet operators, defined in eqs.~(\ref{GIgluOP}) and (\ref{GIqS}), is much more complicated, because they mix with non-physical operators \cite{Gross:1974cs,Georgi:1974wnj}, called {\textit{aliens}}. These must be taken into account in the right hand-side of eq.~(\ref{eq:1}), but a priori it is not known how to construct all required alien operators. An explicit basis, valid at the 2-loop level, was worked out by Dixon and Taylor \cite{Dixon:1974ss}. This result is as the foundation of the work by Hamberg and van Neerven \cite{Hamberg:1991qt}, who managed to correctly determine the gluonic anomalous dimensions at two loops, resolving a series of discrepancies in the previous literature.

Nevertheless the structure of the basis proposed by Dixon and Taylor remained somewhat mysterious and therefore difficult to extend to higher orders. The work of Joglekar and Lee  \cite{Joglekar:1975nu,Joglekar:1976eb,Joglekar:1976pe} characterised\footnote{These results were conjectured earlier in \cite{Kluberg-Stern:1974nmx,Kluberg-Stern:1975ebk}. An alternative proof of the Joglekar-Lee theorem is given in \cite{Henneaux:1993jn}.} the alien operators that can mix with the gauge invariant ones to all perturbative orders. The classification introduces two types of terms: operators that are proportional to the {\textbf{equation of motion}} (\textit{EOM operators}) and operators that {\textbf{vanish under BRST}} transformations ({\textit{BRST-exact operators}}). However, Joglekar and Lee did not provide an explicit basis that can be used to compute the anomalous dimensions via eq.~(\ref{eq:1}).

In conclusion, it is not immediately clear how to extend the OPE method beyond two loops. Recent results in this direction were discussed in detail during this conference \cite{BluemleinLL,TongZhiLL}. Here we present the work \cite{Falcioni:2022fdm}, which provides the general construction of aliens in pure Yang Mills theory. This covers to the most complicated pattern of operator mixing and generalises to the construction of \cite{Dixon:1974ss}, extending it to all order. Explicit bases of alien operators are given through four loops and applied to compute the anomalous dimensions of gluonic operators at $N=2$, $4$, $6$.

In order to proceed, we introduce a scalarised version of the gluonic operator in eq.~(\ref{GIgluOP}). This is obtained by contracting $\O^{(N)}_{g;\mu_1\dots\mu_N}$ with the symmetric traceless projector constructed in terms of a lightlike vector $\Delta_\mu$ (i.e. $\Delta_\mu\Delta^\mu=0$), which gives
\begin{equation}
\label{def:O1}
  \O_1^{(N)} = \O^{(N)}_{g;\mu_1\dots\mu_N}\,\Delta^{\mu_1}\dots\Delta^{\mu_N} = \frac{1}{2} \text{Tr}\left[F_{\rho}\,D^{N-1}\,F^{\rho}\right],
\end{equation}
where we use the notation \cite{Hamberg:1991qt}
\begin{equation}
    F^{a}_{\nu} = F^{a}_{\nu\mu}\Delta^\mu,\quad D = D_\mu\,\Delta^\mu,\quad \partial = \partial_\mu\,\Delta^\mu,\quad A^a=A^a_\mu\,\Delta^\mu,
  \end{equation}
The renormalisation of $\O_1^{(N)}$ requires the introduction of an a priori unknown number of alien operators $\O_{j>1}$, which mix with $\O_1$ via the equation
\begin{equation}
\O^{(N)}_{1}(\mu^2) = Z_{1\,i}^{(N)}(\mu^2)\,\O^{(N),\text{bare}}_{i}.
\end{equation}
In this talk we describe 
\begin{itemize}
    \item the construction of the basis of aliens $\O_{j}^{(N)}$, with $j>1$
    \item the calculation of the physical entry $Z_{1\,1}^{(N)}$ and of the gluonic anomalous dimension $\gamma_{g\,g}^{(N)}$.
\end{itemize}

\section{Theoretical framework}
We begin by introducing the Yang-Mills Lagrangian $\L=\L_0+\L_{\mathrm{GF}+\mathrm{G}}$, where 
\begin{align}
    \mathcal{L}_0 &= -\frac{1}{4}\,F^{a}_{\mu\nu}\,F^{a;\mu\nu}
\end{align}
is invariant under gauge transformations
\begin{equation}
\label{def:gaugetransf}
    \delta_\omega A^a_\mu = D^{ab}_\mu\,\omega^b = \partial_\mu\omega^a + g\,f^{abc}A_\mu^b\omega^c.
\end{equation}
$\L_{\mathrm{GF}+\mathrm{G}}$ comprises gauge fixing and ghost terms, which read
\begin{align}
\label{def:LGF+G}
\mathcal{L}_{\mathrm{GF}+\mathrm{G}} &= s\left[\bar{c}^a\left(\partial^\mu A^a_\mu-\frac{\xi_L}{2}b^a\right)\right] = \bar{s}\left[c^a\left(\frac{\xi_L}{2}b^a-\partial^\mu A^a_\mu\right)\right],
\end{align}
where $\xi_L$ is the gauge-fixing parameter, $c$ and $\bar{c}$ are the ghost and the antighost fields, respectively, $b$ is the Nakanishi-Lautrup field \cite{Nakanishi:1966zz,Lautrup:1967zz} and $\bar{b}^a=-b^a+f^{abc}\bar{c}^bc^c$ \cite{Binosi:2013cea}. Finally, $s$ and $\bar s$ denote the BRST \cite{Becchi:1975nq,Tyutin:1975qk} and the anti-BRST \cite{Curci:1976bt,Ojima:1980da,Baulieu:1981sb} operators, defined as
\begin{align}
      & s\left(A^a_\mu\right)  = D^{ab}_\mu\,c^b,&\bar{s}\left(A^a_\mu\right)  = D^{ab}_\mu\,\bar{c}^b\label{def:BRST1}\\
      &s\left(c^a\right)  = -\frac{g}{2}f^{abc}\,c^b\,c^c,&\bar{s}\left(\bar{c}^a\right)  = -\frac{g}{2}f^{abc}\,\bar{c}^b\,\bar{c}^c\\
      &s\left(\bar{c}^a\right)  = -b^a,&\bar{s}\left(c^a\right)  = -\bar{b}^a\\
      &s\left(b^a\right) = 0,&\bar{s}\left(\bar{b}^a\right) = 0.
\end{align}
The BRST and the anti-BRST operators anticommute and they are both nilpotent, obeying $s^2=\bar{s}^2=\{s,\bar{s}\}=0$. This guarantees that $\L_{\mathrm{GF}+\mathrm{G}}$, and therefore the Yang-Mills Lagrangian as a whole\footnote{Both the BRST and the anti-BRST transformation of the gluon field are special cases of the gauge transformation in eq.~(\ref{def:gaugetransf}), where we identify $\omega^a$ with $c^a$ or with $\bar{c}^a$, respectively. Therefore, $\L_0$ is automatically invariant under BRST and anti-BRST transformations.}, is invariant under BRST and anti-BRST transformations. The BRST invariance of $\L$ has deep consequence for the renormalisability of the theory \cite{Becchi:1975nq,Zinn-Justin:1974ggz}. 

The main point of ref. \cite{Falcioni:2022fdm} is that we can extend the Yang Mills Lagrangian by including the gauge invariant operator defined in eq.~(\ref{def:O1}) and the alien operators that are required to maintain invariance under a generalised version of the BRST transformations. The remaining part of this section shows how to achieve this.

\subsection{Equation of motion operators}
\label{sec:EOM}
In order to generate the alien operators of EOM type, we take the variation of the action
\begin{equation}
    S_0=\int\,\L_0\, d^dx,
\end{equation}
under field redefinitions $A^a_\mu\to A^a_\mu+\mathcal{G}^a_\mu$. These give
\begin{equation}
\label{eq:OEOMgen}
    \O^{(N)}_{\text{EOM}} = \frac{\delta S_0}{\delta A^a_\mu}\,\mathcal{G}^a_\mu(A^{a_1}_{\mu_1},\partial_{\mu_1} A^{a_1}_{\mu_2},\dots) = (D^\nu F_{\nu})^a\,\mathcal{G}^a(A^{a_1},\partial A^{a_1}\dots),
\end{equation}
where we contract the projector $\Delta_{\mu_1}\dots\Delta_{\mu_N}$ to select the symmetric traceless component, as done for the gauge invariant operator in eq.~(\ref{def:O1}), and where we use $\mathcal{G}^a_\mu = \Delta_\mu\,\mathcal{G}^a(A^{a_i},\partial A^{a_i}, \dots)$. The latter is a polynomial in the field $A^a$ and its derivatives
\begin{equation}
\label{eq:ansatzG}
    \mathcal{G}^a(A^{a_i},\partial A^{a_i}, \dots) = \sum_{k=1}^{\infty} g^{k-1}\sum_{\substack{i_1+\dots+i_k\\=N-k-1}}\,C^{a;a_1\dots a_k}_{i_1\dots i_k}\,\left(\partial^{i_1}A^{a_1}\right)...\left(\partial^{i_k}A^{a_k}\right)\,,
\end{equation}
where $C^{a;a_1\dots a_k}_{i_1\dots i_k}$ are coefficients constrained only by colour conservation \cite{Falcioni:2022fdm}. In turn, $\O_{\text{EOM}}^{(N)}$ reads
\begin{align}
\label{eq:EOMgeneral1}
\O_{\text{EOM}}^{(N)}&=\sum_{k=1}^\infty\,\O_{\text{EOM}}^{(N),k}\\
\label{eq:EOMgeneral2}
\O_{\text{EOM}}^{(N),k} &= g^{k-1}\left(D.F\right)^a\sum_{\substack{i_1+\dots+i_k\\=N-k-1}}\,C^{a;a_1\dots a_k}_{i_1\dots i_k}\,\left(\partial^{i_1}A^{a_1}\right)...\left(\partial^{i_k}A^{a_k}\right)\,.
\end{align}

The presence of non-trivial mixing under renormalisation of $\O_1^{(N)}$ into $\O_{\text{EOM}}^{(N)}$ has a transparent diagrammatic interpretation. The insertion of $\O_1^{(N)}$ in a Feynman diagram generates ultraviolet divergences that are proportional to vertices of $\O_{\text{EOM}}^{(N)}$. Up to four-loop order, the OMEs of $\O_1^{(N)}$ feature divergent subgraphs with up to 5 external gluons. Therefore, in the of eq.~(\ref{eq:EOMgeneral1}), we need to take into account only the terms $\O_{\text{EOM}}^{(N),k}$ with $k\leq4$, which feature vertices among 5 gluons or less
\begin{align}
    \label{def:EOMop}
    \O_{\text{EOM}}^{(N),1}&= \eta\;(D.F)^a\;\partial^{N-2}A^a\phantom{\sum_{\substack{i+j=\\N-3}}}\\
    \label{def:EOMop2}
    \O_{\text{EOM}}^{(N),2}&= g(D.F)^a      \sum_{\substack{i+j=\\N-3}}C_{ij}^{abc}(\partial^i A^b)(\partial^j A^c)\\
    \label{def:EOMop3}
    \O_{\text{EOM}}^{(N),3}&= g^2(D.F)^a      \sum_{\substack{i+j+k\\=N-4}} C_{ijk}^{abcd}\,  (\partial^{i} A^b)(\partial^{j} A^c)(\partial^{k} A^d)\\
    \label{def:EOMop4}
    \O_{\text{EOM}}^{(N),4}&= g^3(D.F)^a      \sum_{\substack{i+j+k+l\\=N-5}} C_{ijkl}^{abcde}\,  (\partial^{i} A^b)(\partial^{j} A^c)(\partial^{k} A^d)(\partial^{l} A^e).
\end{align}
The coefficients $C^{a;a_1\dots a_k}_{i_1\dots i_k}$ of eqs.~(\ref{def:EOMop2})-(\ref{def:EOMop4}) are further expanded into free coupling constants $\kappa_{i_1\dots i_k}$ and colour factors. The latter correspond to the colour structures of the $k+1$-point gluonic subdiagrams that appear in the OMEs \cite{Falcioni:2022fdm}. We get
\begin{align}
\label{eq:Cij}
C_{ij}^{abc}&=f^{abc}\kappa_{ij}\\
\label{eq:Cijk}
C_{ijk}^{abcd} &=(ff)^{abcd}\kappa_{ijk}^{(1)}+d_4^{abcd}\kappa_{ijk}^{(2)}+d_{\widehat{4ff}}^{abcd} \kappa_{ijk}^{(3)}\\
\label{eq:Cijkl}
C_{ijkl}^{abcde} &= (fff)^{abcde} \kappa^{(1)}_{ijkl} + d_{4f}^{abcde} \kappa_{ijkl}^{(2)}\,,
\end{align}
where we use the notation
\begin{align}
(ff)^{abcd}&=f^{abe}f^{cde},& (fff)^{abcde}&=f^{abm}f^{mcn}f^{nde},\quad \nn\\
\label{eq:colordefs}
d_{4f}^{abcde}&=d_4^{abcm}f^{mde},& d_{4ff}^{abcd}&=d_4^{abmn}f^{mce}f^{edn}\,,\\
d_{\widehat{4ff}}^{abcd}&=d_{4ff}^{abcd}-\frac{1}{3}C_A\,d_4^{abcd}\,,& \nn
\end{align}
and the symmetrised trace is defined in terms of the group generators in the adjoint representation, $(T_A^a)^{bc}=i\,f^{bac}$, by
\begin{equation}
\label{def:d4abcd}
d_4^{abcd}=\frac{1}{4!}[\mathrm{Tr}(T_A^aT_A^bT_A^cT_A^d)+\text{symmetric permutations}]\,.
\end{equation}
\subsection{Generalisation of Gauge and BRST Invariance}
Since $\O^{(N)}_{\text{EOM}}$ is constructed in terms of the gauge field and its derivatives, it manifestly breaks gauge symmetry. However, it maintains the invariance under a generalised type of gauge transformations. To be concrete, we extend the Yang Mills Lagrangian, by including the gauge invariant operator $\O_1$ and the EOM alien operators
\begin{equation}
    \mathcal{L}_{\text{EGI}}=\mathcal{L}_0 + \mathcal{C}_1\,\O_1^{(N)} + \O^{(N)}_{\text{EOM}}, \quad S_{\text{EGI}} = \int \L_{\text{EGI}} d^dx,
\end{equation}
where $\mathcal{C}_1$ is the coupling constant associated to $\O_1^{(N)}$. We consider the variation of $S_{\text{EGI}}$ under 
\begin{equation}
\label{Aomegageneralised}
A^a_\mu \to A^a_\mu + \delta_\omega A^a_\mu + \delta_\omega^\Delta A^a_\mu\,,
\end{equation}
where $\delta_\omega^\Delta A^a_\mu$ is multi-linear in $\Delta$. By working to leading order\footnote{This is sufficient to renormalise a {\textit{single}} insertion of $\O_1^{(N)}$ in any correlator.}in $\Delta$, we determine $\delta_\omega^\Delta A^a_\mu$ such that it cancels the gauge variation of $\O_{\text{EOM}}^{(N)}$ and $\delta S_{\text{EGI}}=0$. This leads to the relation
\begin{equation}
\label{eq:GenGaugeTrans}
  \delta^\Delta_\omega A^a_\mu= -\delta_\omega \mathcal{G}^a_\mu + g\,f^{abc}\,\mathcal{G}^b_\mu\, \omega^c,
\end{equation}
where $\delta_\omega \mathcal{G}^a_\mu$ is the gauge variation of $\mathcal{G}^a_\mu$.
The concrete expression of $\delta_\omega^\Delta A^a_\mu$ is worked out easily by using the ansatz in eq.~(\ref{eq:ansatzG}). However, we notice that eq.~(\ref{eq:GenGaugeTrans}) holds beyond the leading-twist case, upon using a more general ansatz for $\mathcal{G}^a_\mu$ without symmetric traceless projection.

Eq.~(\ref{eq:GenGaugeTrans}) immediately leads to a generalised BRST transformation for the gauge field
\begin{equation}
\label{AgeneralisedBRST}
    A^a_\mu \to s'(A^a_\mu) = s(A^a_\mu) + s_\Delta(A^a_\mu),
\end{equation}
where $s_\Delta(A^a_\mu)$ is found from $\delta_\omega^\Delta A^a_\mu$, in eq.~(\ref{eq:GenGaugeTrans}), by replacing $\omega^a\to c^a$
\begin{equation}
\label{eq:GenBRSTtrans}
 s_\Delta (A^a_\mu) = -s\left(\mathcal{G}^a_\mu\right) + g\,f^{abc}\,\mathcal{G}^b_\mu\,c^c.
\end{equation}
Here $s\left(\mathcal{G}^a_\mu\right)$ indicates the usual BRST operator, defined in eq.~(\ref{def:BRST1}), acting on $\mathcal{G}^a_\mu$. The crucial feature of eq.~(\ref{eq:GenBRSTtrans}) is that, for any choice of $\mathcal{G}^a_\mu$, the generalised BRST transformation is nilpotent
\begin{equation}
  s'^2(A^a_\mu) = \{s,s_\Delta\}\,A^a_\mu = 0.
\end{equation}
This provides the missing piece to construct a Lagrangian that is invariant under generalised BRST transformations and includes $\O_1^{(N)}$ and the EOM alien operators $\O_{\text{EOM}}^{(N)}$
\begin{equation}
\label{eq:Ltilde}
  \widetilde{\mathcal{L}} = \mathcal{L}_{\text{EGI}}+s'\left[\bar{c}^a\left(\partial^\mu A^a_\mu-\frac{\xi_L}{2}b^a\right)\right],
\end{equation}
where we assume $s_\Delta(b^a) = s_\Delta(\bar{c}) = 0$. By construction, $\L_{\text{EGI}}$ is invariant under the transformation in eq.~(\ref{AgeneralisedBRST}), because the latter has the same form of eq.~(\ref{Aomegageneralised}). The second term in eq.~(\ref{eq:Ltilde}), which is constructed to recover the gauge fixing and ghost term of eq.~(\ref{def:LGF+G}), is BRST-exact (now in the generalised sense), thus ensuring invariance of $\widetilde{\L}$.
\subsection{BRST-exact operators and summary}
In the last part of this section we read off the BRST-exact alien operators from eq.~(\ref{eq:Ltilde})
\begin{equation}
\label{eq:OGgeneral}
    \O^{(N)}_{G} = s_{\Delta}\left(\bar{c}^a\,\partial^\mu A^a_\mu\right)
    = 
\bar{c}^a\partial^\mu\left(s\left(\mathcal{G}^a_\mu\right) - g\,f^{abc}\,\mathcal{G}^b_\mu\,c^c\right),
\end{equation}
where we used eq.~(\ref{eq:GenBRSTtrans}) and $s_\Delta(\bar{c}^a)=0$. For every value of $N$, we get explicit expression by replacing eq.~(\ref{eq:ansatzG}) in the equation above and operating with the operator $s$. For instance, for $N=2$ we have $\mathcal{G}^a = \eta A^a$, giving
\begin{align}
    O_{\text{EOM}}^{(2)} = \eta\,(D.F)^a\,A^a,\qquad O_{G}^{(2)} = \eta\,\bar{c}^a\partial^2 c^a.    
\end{align}
Increasing $N$, more and more terms in $\mathcal{G}^a_\mu$, eq.~(\ref{eq:ansatzG}), contribute. For instance, for $N=4$ we get
\begin{equation}
\label{eq:G4}
    \mathcal{G}^a = \eta\;\partial^2 A^a + 2\,g\kappa_{01}\,f^{aa_1a_2}\,A^{a_1}\partial A^{a_2} + g^2\,\kappa_{000}^{(2)}\,d^{aa_1a_2a_3} A^{a_1}A^{a_2}A^{a_3}.
\end{equation}
By plugging this expression in eq.~(\ref{eq:OGgeneral}) we get
\begin{align}
\label{eq:O4G}
    \O^{(4)}_G= &-\eta\,\partial\bar{c}^a\Big[\partial^3 c^a+g\,f^{abc}\Big(2\,\partial A^{b}\partial c^{c}+A^{b}\partial^2c^{c}\Big)\Big]\\
    &-2\,g\kappa_{01}\partial\bar{c}^a\Big[f^{abc}\Big(A^{b}\partial^2c^{c}-\partial A^{b}\partial c^{c}\Big)+g\,f^{abz}f^{cdz}\,A^{b}A^{c}\partial c^{d}\Big]\nn\\
    &-3g^2\,\kappa_{000}^{(2)}d_4^{abcd}\,\partial\bar{c}^aA^bA^c\partial c^d\nn
  \end{align}
However, not all the terms above are independent. A set of non-trivial relations on the coefficient $\kappa_{i_1\dots i_k}$ derives by imposing the invariance of $\widetilde{L}$ under generalised anti-BRST. We already pointed out that the Yang-Mills Lagrangian can be written as the variation of an {\textit{ancestor}} operator under anti-BRST transformations. The latter are defined to be identical to BRST transformation, where the ghost field is exchanged with an antighost. In the same way, we construct generalised anti-BRST transformations from eq.~(\ref{eq:GenBRSTtrans}) 
\begin{equation}
\label{eq:GenAntiBRSTtrans}
    \bar{s}'A^a_\mu = \bar{s} A^a_\mu + \bar{s}_\Delta A^a_\mu,\qquad \bar{s}_\Delta A^a_\mu = -\bar{s}\left(\mathcal{G}^a_\mu\right) + g\,f^{abc}\,\mathcal{G}^b_\mu\,\bar{c}^c,
\end{equation}
and $\bar{s}_\Delta c^a =\bar{s}_{\Delta} b^a = 0$. By using the generalised anti-BRST in the gauge fixing and ghost Lagrangian, eq.~(\ref{def:LGF+G}), we derive the following expression for $\widetilde{\L}$
\begin{align}
\label{eq:antiBRSTLtilde}
\widetilde{\mathcal{L}}=\mathcal{L}_0\,+\,\mathcal{C}_1\,\O^{(N)}_1+\O^{(N)}_{\text{EOM}} \;+\;\bar{s}'\left[c^a\left(\frac{\xi_L}{2}b^a-\partial^\mu A^a_\mu\right)\right].
\end{align}
The equivalence of the right hand-sides of eqs.~(\ref{eq:Ltilde}) and (\ref{eq:antiBRSTLtilde}) implies
\begin{equation}
\label{eq:BRSTvsAntiBRST}
  -\bar{c}^a\,\partial^\mu\left(s_\Delta A^a_\mu\right) = c^a\,\partial^\mu\left(\bar{s}_\Delta A^a_\mu\right),
\end{equation}
which relates different couplings in $\mathcal{G}^a$ \cite{Falcioni:2022fdm}. For instance, by defining $s_\Delta$ and $\bar{s}_{\Delta}$ as in eqs.~(\ref{eq:GenBRSTtrans}) and (\ref{eq:GenAntiBRSTtrans}), respectively, where $\mathcal{G}^a$ is given in eq.~(\ref{eq:G4}) for $N=4$, we find that $\kappa_{01}=\frac{\eta}{2}$, thus reducing eq.~(\ref{eq:O4G}) to two independent terms.

To summarise, we construct the Lagrangian that includes all the alien operators associated to $\O_1^{(N)}$ by requiring its invariance under a generalised type of BRST (and anti-BRST) transformation, which gives
\begin{equation}
\label{Lfinal}
\widetilde{\L} = \L_0-\frac{\left(\partial^\mu A^a_\mu\right)^2}{2\xi_L} -\bar{c}^a\partial^\mu D^{ab}_\mu c^b + \mathcal{C}_1\;\O^{(N)}_1 + \sum_{i>1}\mathcal{C}_i\,\O^{(N)}_i, 
\end{equation}
where the {\textit{alien}} sector is
\begin{equation}
    \sum_{i>1}\mathcal{C}_i\,\O^{(N)}_i = \O_{\text{EOM}}^{(N)}+\O_G^{(N)}.
\end{equation}
Both $\O_{\text{EOM}}^{(N)}$ and $\O_G^{(N)}$ are computed in terms of $\mathcal{G}^a$ in the form of eq.~(\ref{eq:ansatzG}), as follows 
\begin{itemize}
    \item $\O^{(N)}_{\text{EOM}} = (D^\mu F_\mu)^a\,\mathcal{G}^a$
    \item $\O^{(N)}_{G} = -(\partial\bar{c}^a)\,\left(s\left(\mathcal{G}^a\right) - g\,f^{abc}\,\mathcal{G}^b\,c^c\right)$
\end{itemize}
We reduce to a minimal set of operators by imposing the anti-BRST invariance, given in eq.~(\ref{eq:BRSTvsAntiBRST}). This identifies the minimal set of coupling constants $\mathcal{C}_i$, with $i>1$, in eq.~(\ref{Lfinal}).
\section{Calculations and results}
We employ the bases of alien operators defined in the previous section to compute the physical anomalous dimension $\gamma_{1\,1}^{(N)}$ with off-shell OMEs. Following \cite{Falcioni:2022fdm}, we work in the background field method \cite{DeWitt:1967ub,Honerkamp:1971sh,Honerkamp:1972fd,Kallosh:1974yh,Arefeva:1974jv,Sarkar:1974db,Sarkar:1974ni,Kluberg-Stern:1974nmx,tHooft:1975uxh,Grisaru:1975ei,Abbott:1980hw,Abbott:1981ke} and we define the OMEs with two external background fields and an insertion of the operator $\O_i$
\begin{equation}
  \label{eq:GammaOBBmunu}
  \left(\Gamma_{\O_i;BB}^{(N)}\right)^{a_1 a_2}_{\nu_1\nu_2}(g,\xi;p^2) = \int d^dx_1 d^dx_2\,e^{ip \cdot( x_1-x_2)}\,\langle 0 | T\left[B^{a_1}_{\nu_1}(x_1)B^{a_2}_{\nu_2}(x_2)\O_i^{(N)}(0)\right]|0\rangle_{\mathrm{1PI}},
\end{equation}
where $\xi=1-\xi_L$ and $B^a_\mu$ is the background field. These quantities vanish at tree level unless the inserted operator is the physical one, namely $\O_1^{(N)}$, and we have\footnote{
We can reduce to scalar quantities by applying projectors on the colour and spin indices, e.g.
\begin{equation}
 \label{eq:GammaOBB}
 \Gamma_{i;BB}^{(N)}(g_B,\xi_B,p^2) = \frac{\delta^{a_1a_2}}{N_A}\;\frac{g^{\nu_1\nu_2}}{(d-1)}\;\left(\Gamma_{i;BB}^{(N)}\right)^{a_1 a_2}_{\nu_1\nu_2}(g_B,\xi_B,p^2),
\end{equation}}
\begin{equation}
    \raisebox{17.5pt}{$\Gamma_{i;BB}^{(N)}(g,\xi,p^2)\,=\,$}
    \begin{axopicture}(60,40)
      \DoubleGluon(0,20)(15,20){1}{3}{1}
      \DoubleGluon(45,20)(60,20){1}{3}{1}
      \GCirc(30,20){15}{.8}
      \BCirc(30,35){4}
      \Line(28,33)(32,37)
      \Line(28,37)(32,33)
      {\SetPFont{}{8}{
      \PText(28,39)(0)[rb]{$i$}}}
    \end{axopicture}
    \raisebox{17.5pt}{$\displaystyle \,=\,\left\{
      \begin{array}{lr}
        \Gamma^{(N),0}_{i;BB} + \delta\Gamma_{i;BB}^{(N)}(g,\xi,p^2) & i=1\\\\
        \delta \Gamma_{i;BB}^{(N)}(g,\xi,p^2) & i\neq 1
      \end{array}
      \right.$}
\end{equation}
The anomalous dimensions $\gamma_{1\,1}^{(N)}$, for each value of $N$, are computed in terms of the renormalisation constants $Z_{1\,1}^{(N)} = 1+Z_{1\,1}^{(N)}$ via
\begin{equation}
    \gamma_{\mathrm{g}\,\mathrm{g}}^{(N)} = a\frac{\partial}{\partial a}Z_{1\,1}^{(N)}\left|_{\frac{1}{\epsilon}}\right.,
\end{equation}
where $a=g^2/(16\pi^2)$ is the strong coupling constant and $\epsilon=(4-d)/2$ is the dimensional regularisation parameter. We extract $Z_{1\,1}$ from the off-shell OMEs by solving
\begin{equation}
\label{eq:deltaZ11}
    \delta Z_{1\,1}^{(N)} = -\frac{1}{Z_B\,\Gamma_{1;BB}^{(N),0}}K_\epsilon\bigg[Z_B\sum_{i\geq1}Z^{(N)}_{1\,i}\delta\Gamma^{(N)}_{i;BB}(g_B,\xi_B)\bigg],
\end{equation}
where the operator $K_\epsilon$ extracts the poles in the Laurent expansion in $\epsilon$ and where
the renormalisation constant $Z_B$ of the background field is related to the renormalisation of the coupling by $Z_B\,Z_a=1$ \cite{Abbott:1980hw,Abbott:1981ke}. The constants $Z_{1\,i>1}^{(N)}$, which generate mixing between $\O_1^{(N)}$ and the alien operators, are computed in \cite{Falcioni:2022fdm} by applying the $R^*$ operation to appropriate correlators\footnote{See \cite{deVries:2019nsu} for a discussion of the method}, which feature the insertion of $\O_1^{(N)}$ and a pair ghost-antighost and gluons as external lines. 

Beginning with $N=2$, in addition to the gauge invariant operator $\O_1^{(2)}$ we find only one alien, leading to the basis
\begin{equation*}
    \O_1^{(2)}=\frac{1}{2} F^a_\mu\,F^{a;\mu},\qquad \O_2^{(2)} = (D^\nu F_\nu)^a\,A^a + \bar{c}^a\partial^2c^a.
\end{equation*}
After computing the mixing renormalisation constant up to 3-loop  
\begin{align}
\delta Z_{1\,2}^{(2)}&=-a\frac{C_A}{2\epsilon}+a^2\,C_A^2\left[\frac{19}{24\epsilon^2}+\frac{5}{48}\frac{\xi}{\epsilon}-\frac{35}{48\epsilon}\right]+a^3\,C_A^3\Big[ - \frac{779}{432\epsilon^3}+\frac{1}{\epsilon^2}\Big(\frac{2807}{864}-\frac{35\xi}{216}+\frac{5\xi^2}{288}\Big)\nn\\
&+\frac{1}{\epsilon}\Big(-\frac{16759}{7776}-\frac{11\zeta_3}{72}+\frac{377\xi}{1728}+\frac{5\zeta_3\,\xi}{72}-\frac{65\xi^2}{1728}\Big)\Big]+O(a^4),
\end{align}
we find
\begin{equation}
    \delta Z_{1\,1}^{(2)} = 0,    
\end{equation}
which agrees with the recent result of an explicit calculation in a different approach \cite{Moch:2021qrk} and with general the theorem proved in \cite{Freedman:1974gs,Freedman:1974ze}.

The basis of operators at $N=4$ involves two aliens, as discussed in the previous section. One possible choice is  
\begin{align}
  \O_1^{(4)}&=\frac{1}{2}\text{Tr}\big[F_{\nu}D^2F^{\nu}\big],\\
  \O_2^{(4)}&=(D.F)^a\Big[ \partial^2A^a + gf^{abc}A^{b}\partial A^{c}\Big]-\partial \bar{c}^a\,\partial^3c^a - gf^{abc}\,\partial \bar{c}^a\Big[2A^{b}\partial^2 c^{c}+\partial A^{b}\,\partial c^{c}\Big]\nn\\
  &-g^2\,f^{abe}f^{cde}\,\partial \bar{c}^a\,A^{b}A^{c}\,\partial c^{d},\\
  \O_3^{(4)}&=d^{abcd}\Big[(D.F)^aA^{b}A^{c}A^{d}-3\partial \bar{c}^a\,A^{b}A^{c}\,\partial c^{d}\Big].
\end{align}
We notice that $\O_2^{(4)}$ generates a vertex between 2 gluon or 2 ghost lines, and therefore it can enter the 4-loop correlator $\Gamma_{1;BB}^{(4)}$ as the counterterm of a two-point subdiagram with at most 3 loops. In contrast, $\O_3^{(4)}$ involves vertices with at least 4 external particles. These can enter as subdiagrams with at most one loop. As a consequence the mixing renormalisation constants $\delta Z_{1\,2}^{(4)}$ and $\delta Z_{1\,3}^{(4)}$ are computed up to three and one loop, respectively
\begin{align}
    \delta Z_{1\,2}^{(4)}&=-\frac{aC_A}{12\epsilon}-a^2C_A^2\big[\frac{97}{1440\epsilon^2}-\frac{\xi}{320\epsilon}+\frac{8641}{86400\epsilon}\big]+a^3C_A^3\Big[\frac{9437}{86400\epsilon^3}-\frac{1}{\epsilon^2}\Big(\frac{1520341}{15552000}-\frac{853\xi}{86400}\Big)\nn\\
      &-\frac{1}{\epsilon}\Big(\frac{166178237}{466560000}+\frac{\zeta_3}{2400}-\frac{37199\xi}{648000}-\frac{37\zeta_3\,\xi}{9600}\Big)\Big]\\
    \delta Z_{1\,3}^{(4)}&=\frac{a\,C_A}{24\epsilon}.
  \end{align}
  Using these results in eq.~(\ref{eq:deltaZ11}) we get
  \begin{align}
        \delta Z^{(4)}_{1\,1}&= a\frac{21C_A}{5\epsilon}\!+\!a^2C_A^2\!\left(\frac{28}{25\epsilon^2}+\frac{7121}{1000\epsilon}\right)\!
        -\!a^3C_A^3\!\left(\frac{1316}{1125\epsilon^3}+\frac{151441}{45000\epsilon^2}-\frac{103309639}{4050000\epsilon}\right)\nn\\
        &\hspace{-.2cm}+ a^4\left\{C_A^4\left[\frac{11186}{5625\epsilon^4}+ \frac{1512989}{450000\epsilon^3}-\frac{5437269017}{162000000\epsilon^2}+\frac{1}{\epsilon}\Big(\frac{1502628149}{13500000} + \frac{1146397\zeta_3}{45000}\right.\right.\nn\\
        &\left.\left.\hspace{-.2cm}-\frac{126\zeta_5}{5}\Big)\right]  + \frac{d_{AA}}{N_A}\,\left(\frac{21623}{600\epsilon}+\frac{3899\,\zeta_3}{15\epsilon}-\frac{1512\,\zeta_5}{5\epsilon}\right)\right\},
\end{align}
which agrees with the state-of-the-art result \cite{Moch:2021qrk}.      

In the paper \cite{Falcioni:2022fdm}, we also discuss the renormalisation of $\O_1^{(6)}$ up to three loops, finding agreement with the work \cite{Nogueira:1991ex}, which relies on a different method. From the computational point of view, getting the four-loop anomalous dimension of $\O_1^{(6)}$ is challenging. The bottleneck is the determination of the mixing renormalisation constant, which relies on $R^*$ in our current implementation. The computational cost of this technique increases with insertion of operators of high mass dimension. Ultimately, we might employ different methods for this part of the calculation, e.g. the one discussed in \cite{Misiak:1994zw}.

\section{Conclusion}
While the OPE framework is a powerful tool to compute the four-loop splitting functions from the anomalous dimensions of gauge invariant operators, in the flavour sector the mixing of the gauge invariant operators and unphysical {\textit{aliens}} poses a severe conceptual issue. Indeed, it was not known how to construct the required alien operators beyond two loops. Here we discuss a general procedure to construct all the aliens that mix with the gluonic operator, $\O_1^{(N)}$. This procedure leverages the generalised BRST transformation of eq.~(\ref{eq:GenBRSTtrans}) to construct the EOM and the BRST-exact operators, given respectively in eqs.~(\ref{eq:OEOMgen}) and (\ref{eq:OGgeneral}). The concrete expression of the transformation is given up to 4 loops and can be read off eqs.~(\ref{def:EOMop})-(\ref{def:EOMop4}). We verify the approach by renormalising the operators $\O_1^{(2)}$ and $\O_1^{(4)}$ through 4 loops and $\O_1^{(6)}$ through 3 loops.

In the future, in addition to technical work on the extraction of mixing renormalisation constants, we plan to extend this method to fermionic contributions. This will allow us to complete the theoretical framework to compute the whole flavour singlet sector of the splitting functions up to four-loop order.

\section*{Acknowledgements}\noindent
I would like to thank Franz Herzog for his invaluable collaboration. I was supported by the ERC Starting Grant 715049 "QCDforfuture" with Principal Investigator Jennifer
Smillie and by the STFC Consolidated Grant "Particle Physics at the Higgs Centre". 
\bibliographystyle{JHEP}
\bibliography{refs}

\providecommand{\href}[2]{#2}\begingroup\raggedright\begin{thebibliography}{10}

\bibitem{Anastasiou:2015vya}
C.~Anastasiou, C.~Duhr, F.~Dulat, F.~Herzog and B.~Mistlberger, \emph{{Higgs
  Boson Gluon-Fusion Production in QCD at Three Loops}},
  \href{http://dx.doi.org/10.1103/PhysRevLett.114.212001}{\emph{Phys. Rev.
  Lett.} {\bf 114} (2015) 212001} [\href{https://arxiv.org/abs/1503.06056}{{\tt
  arXiv:1503.06056}}].

\bibitem{Anastasiou:2016cez}
C.~Anastasiou, C.~Duhr, F.~Dulat, E.~Furlan, T.~Gehrmann, F.~Herzog et~al.,
  \emph{{High precision determination of the gluon fusion Higgs boson
  cross-section at the LHC}},
  \href{http://dx.doi.org/10.1007/JHEP05(2016)058}{\emph{JHEP} {\bf 05} (2016)
  058} [\href{https://arxiv.org/abs/1602.00695}{{\tt arXiv:1602.00695}}].

\bibitem{Mistlberger:2018etf}
B.~Mistlberger, \emph{{Higgs boson production at hadron colliders at N$^{3}$LO
  in QCD}}, \href{http://dx.doi.org/10.1007/JHEP05(2018)028}{\emph{JHEP} {\bf
  05} (2018) 028} [\href{https://arxiv.org/abs/1802.00833}{{\tt
  arXiv:1802.00833}}].

\bibitem{Duhr:2019kwi}
C.~Duhr, F.~Dulat and B.~Mistlberger, \emph{{Higgs Boson Production in
  Bottom-Quark Fusion to Third Order in the Strong Coupling}},
  \href{http://dx.doi.org/10.1103/PhysRevLett.125.051804}{\emph{Phys. Rev.
  Lett.} {\bf 125} (2020) 051804} [\href{https://arxiv.org/abs/1904.09990}{{\tt
  arXiv:1904.09990}}].

\bibitem{Duhr:2020seh}
C.~Duhr, F.~Dulat and B.~Mistlberger, \emph{{Drell-Yan Cross Section to Third
  Order in the Strong Coupling Constant}},
  \href{http://dx.doi.org/10.1103/PhysRevLett.125.172001}{\emph{Phys. Rev.
  Lett.} {\bf 125} (2020) 172001} [\href{https://arxiv.org/abs/2001.07717}{{\tt
  arXiv:2001.07717}}].

\bibitem{Chen:2021isd}
X.~Chen, T.~Gehrmann, E.W.N.~Glover, A.~Huss, B.~Mistlberger and A.~Pelloni,
  \emph{{Fully Differential Higgs Boson Production to Third Order in QCD}},
  \href{http://dx.doi.org/10.1103/PhysRevLett.127.072002}{\emph{Phys. Rev.
  Lett.} {\bf 127} (2021) 072002} [\href{https://arxiv.org/abs/2102.07607}{{\tt
  arXiv:2102.07607}}].

\bibitem{Duhr:2020sdp}
C.~Duhr, F.~Dulat and B.~Mistlberger, \emph{{Charged current Drell-Yan
  production at N$^{3}$LO}},
  \href{http://dx.doi.org/10.1007/JHEP11(2020)143}{\emph{JHEP} {\bf 11} (2020)
  143} [\href{https://arxiv.org/abs/2007.13313}{{\tt arXiv:2007.13313}}].

\bibitem{Floratos:1978ny}
E.G.~Floratos, D.A.~Ross and C.T.~Sachrajda, \emph{{Higher Order Effects in
  Asymptotically Free Gauge Theories. 2. Flavor Singlet Wilson Operators and
  Coefficient Functions}},
  \href{http://dx.doi.org/10.1016/0550-3213(79)90094-4}{\emph{Nucl. Phys. B}
  {\bf 152} (1979) 493}.

\bibitem{Gonzalez-Arroyo:1979qht}
A.~Gonzalez-Arroyo and C.~Lopez, \emph{{Second Order Contributions to the
  Structure Functions in Deep Inelastic Scattering. 3. The Singlet Case}},
  \href{http://dx.doi.org/10.1016/0550-3213(80)90207-2}{\emph{Nucl. Phys. B}
  {\bf 166} (1980) 429}.

\bibitem{Furmanski:1980cm}
W.~Furmanski and R.~Petronzio, \emph{{Singlet Parton Densities Beyond Leading
  Order}}, \href{http://dx.doi.org/10.1016/0370-2693(80)90636-X}{\emph{Phys.
  Lett. B} {\bf 97} (1980) 437}.

\bibitem{Hamberg:1991qt}
R.~Hamberg and W.L.~van~Neerven, \emph{{The Correct renormalization of the
  gluon operator in a covariant gauge}},
  \href{http://dx.doi.org/10.1016/0550-3213(92)90593-Z}{\emph{Nucl. Phys. B}
  {\bf 379} (1992) 143}.

\bibitem{Vogelsang:1995vh}
W.~Vogelsang, \emph{{A Rederivation of the spin dependent next-to-leading order
  splitting functions}},
  \href{http://dx.doi.org/10.1103/PhysRevD.54.2023}{\emph{Phys. Rev. D} {\bf
  54} (1996) 2023} [\href{https://arxiv.org/abs/hep-ph/9512218}{{\tt
  hep-ph/9512218}}].

\bibitem{Mertig:1995ny}
R.~Mertig and W.L.~van~Neerven, \emph{{The Calculation of the two loop spin
  splitting functions P(ij)(1)(x)}},
  \href{http://dx.doi.org/10.1007/s002880050138}{\emph{Z. Phys. C} {\bf 70}
  (1996) 637} [\href{https://arxiv.org/abs/hep-ph/9506451}{{\tt
  hep-ph/9506451}}].

\bibitem{Ellis:1996nn}
R.K.~Ellis and W.~Vogelsang, \emph{{The Evolution of parton distributions
  beyond leading order: The Singlet case}},
  \href{https://arxiv.org/abs/hep-ph/9602356}{{\tt hep-ph/9602356}}.

\bibitem{Matiounine:1998ky}
Y.~Matiounine, J.~Smith and W.L.~van~Neerven, \emph{{Two loop operator matrix
  elements calculated up to finite terms}},
  \href{http://dx.doi.org/10.1103/PhysRevD.57.6701}{\emph{Phys. Rev. D} {\bf
  57} (1998) 6701} [\href{https://arxiv.org/abs/hep-ph/9801224}{{\tt
  hep-ph/9801224}}].

\bibitem{Matiounine:1998re}
Y.~Matiounine, J.~Smith and W.L.~van~Neerven, \emph{{Two loop operator matrix
  elements calculated up to finite terms for polarized deep inelastic lepton -
  hadron scattering}},
  \href{http://dx.doi.org/10.1103/PhysRevD.58.076002}{\emph{Phys. Rev. D} {\bf
  58} (1998) 076002} [\href{https://arxiv.org/abs/hep-ph/9803439}{{\tt
  hep-ph/9803439}}].

\bibitem{Larin:1993vu}
S.A.~Larin, T.~van~Ritbergen and J.A.M.~Vermaseren, \emph{{The Next
  next-to-leading QCD approximation for nonsinglet moments of deep inelastic
  structure functions}},
  \href{http://dx.doi.org/10.1016/0550-3213(94)90268-2}{\emph{Nucl. Phys. B}
  {\bf 427} (1994) 41}.

\bibitem{Larin:1996wd}
S.A.~Larin, P.~Nogueira, T.~van~Ritbergen and J.A.M.~Vermaseren, \emph{{The
  Three loop QCD calculation of the moments of deep inelastic structure
  functions}},
  \href{http://dx.doi.org/10.1016/S0550-3213(97)80038-7}{\emph{Nucl. Phys. B}
  {\bf 492} (1997) 338} [\href{https://arxiv.org/abs/hep-ph/9605317}{{\tt
  hep-ph/9605317}}].

\bibitem{Moch:2004pa}
S.~Moch, J.A.M.~Vermaseren and A.~Vogt, \emph{{The Three loop splitting
  functions in QCD: The Nonsinglet case}},
  \href{http://dx.doi.org/10.1016/j.nuclphysb.2004.03.030}{\emph{Nucl. Phys. B}
  {\bf 688} (2004) 101} [\href{https://arxiv.org/abs/hep-ph/0403192}{{\tt
  hep-ph/0403192}}].

\bibitem{Vogt:2004mw}
A.~Vogt, S.~Moch and J.A.M.~Vermaseren, \emph{{The Three-loop splitting
  functions in QCD: The Singlet case}},
  \href{http://dx.doi.org/10.1016/j.nuclphysb.2004.04.024}{\emph{Nucl. Phys. B}
  {\bf 691} (2004) 129} [\href{https://arxiv.org/abs/hep-ph/0404111}{{\tt
  hep-ph/0404111}}].

\bibitem{Ablinger:2014nga}
J.~Ablinger, A.~Behring, J.~Bl\"umlein, A.~De~Freitas, A.~von~Manteuffel and
  C.~Schneider, \emph{{The 3-loop pure singlet heavy flavor contributions to
  the structure function $F_2(x,Q^2)$ and the anomalous dimension}},
  \href{http://dx.doi.org/10.1016/j.nuclphysb.2014.10.008}{\emph{Nucl. Phys. B}
  {\bf 890} (2014) 48} [\href{https://arxiv.org/abs/1409.1135}{{\tt
  arXiv:1409.1135}}].

\bibitem{Ablinger:2017tan}
J.~Ablinger, A.~Behring, J.~Bl\"umlein, A.~De~Freitas, A.~von~Manteuffel and
  C.~Schneider, \emph{{The three-loop splitting functions $P_{qg}^{(2)}$ and
  $P_{gg}^{(2, N_F)}$}},
  \href{http://dx.doi.org/10.1016/j.nuclphysb.2017.06.004}{\emph{Nucl. Phys. B}
  {\bf 922} (2017) 1} [\href{https://arxiv.org/abs/1705.01508}{{\tt
  arXiv:1705.01508}}].

\bibitem{Behring:2019tus}
A.~Behring, J.~Bl\"umlein, A.~De~Freitas, A.~Goedicke, S.~Klein,
  A.~von~Manteuffel et~al., \emph{{The Polarized Three-Loop Anomalous
  Dimensions from On-Shell Massive Operator Matrix Elements}},
  \href{http://dx.doi.org/10.1016/j.nuclphysb.2019.114753}{\emph{Nucl. Phys. B}
  {\bf 948} (2019) 114753} [\href{https://arxiv.org/abs/1908.03779}{{\tt
  arXiv:1908.03779}}].

\bibitem{Ablinger:2019etw}
J.~Ablinger, A.~Behring, J.~Bl\"umlein, A.~De~Freitas, A.~von~Manteuffel,
  C.~Schneider et~al., \emph{{The three-loop single mass polarized pure singlet
  operator matrix element}},
  \href{http://dx.doi.org/10.1016/j.nuclphysb.2020.114945}{\emph{Nucl. Phys. B}
  {\bf 953} (2020) 114945} [\href{https://arxiv.org/abs/1912.02536}{{\tt
  arXiv:1912.02536}}].

\bibitem{Blumlein:2021enk}
J.~Bl\"umlein, P.~Marquard, C.~Schneider and K.~Sch\"onwald, \emph{{The
  three-loop unpolarized and polarized non-singlet anomalous dimensions from
  off shell operator matrix elements}},
  \href{http://dx.doi.org/10.1016/j.nuclphysb.2021.115542}{\emph{Nucl. Phys. B}
  {\bf 971} (2021) 115542} [\href{https://arxiv.org/abs/2107.06267}{{\tt
  arXiv:2107.06267}}].

\bibitem{Blumlein:2021ryt}
J.~Bl\"umlein, P.~Marquard, C.~Schneider and K.~Sch\"onwald, \emph{{The
  three-loop polarized singlet anomalous dimensions from off-shell operator
  matrix elements}},
  \href{http://dx.doi.org/10.1007/JHEP01(2022)193}{\emph{JHEP} {\bf 01} (2022)
  193} [\href{https://arxiv.org/abs/2111.12401}{{\tt arXiv:2111.12401}}].

\bibitem{Gross:1974cs}
D.J.~Gross and F.~Wilczek, \emph{{Asymptotically free gauge theories. 2.}},
  \href{http://dx.doi.org/10.1103/PhysRevD.9.980}{\emph{Phys. Rev. D} {\bf 9}
  (1974) 980}.

\bibitem{Georgi:1974wnj}
H.~Georgi and H.D.~Politzer, \emph{{Electroproduction scaling in an
  asymptotically free theory of strong interactions}},
  \href{http://dx.doi.org/10.1103/PhysRevD.9.416}{\emph{Phys. Rev. D} {\bf 9}
  (1974) 416}.

\bibitem{Gracey:1996ad}
J.A.~Gracey, \emph{{Anomalous dimensions of operators in polarized deep
  inelastic scattering at O(1/N(f))}},
  \href{http://dx.doi.org/10.1016/S0550-3213(96)00485-3}{\emph{Nucl. Phys. B}
  {\bf 480} (1996) 73} [\href{https://arxiv.org/abs/hep-ph/9609301}{{\tt
  hep-ph/9609301}}].

\bibitem{Davies:2016jie}
J.~Davies, A.~Vogt, B.~Ruijl, T.~Ueda and J.A.M.~Vermaseren, \emph{{Large-$n_f$
  contributions to the four-loop splitting functions in QCD}},
  \href{http://dx.doi.org/10.1016/j.nuclphysb.2016.12.012}{\emph{Nucl. Phys. B}
  {\bf 915} (2017) 335} [\href{https://arxiv.org/abs/1610.07477}{{\tt
  arXiv:1610.07477}}].

\bibitem{Moch:2017uml}
S.~Moch, B.~Ruijl, T.~Ueda, J.A.M.~Vermaseren and A.~Vogt, \emph{{Four-Loop
  Non-Singlet Splitting Functions in the Planar Limit and Beyond}},
  \href{http://dx.doi.org/10.1007/JHEP10(2017)041}{\emph{JHEP} {\bf 10} (2017)
  041} [\href{https://arxiv.org/abs/1707.08315}{{\tt arXiv:1707.08315}}].

\bibitem{Ruijl:2017cxj}
B.~Ruijl, T.~Ueda and J.A.M.~Vermaseren, \emph{{Forcer, a FORM program for the
  parametric reduction of four-loop massless propagator diagrams}},
  \href{http://dx.doi.org/10.1016/j.cpc.2020.107198}{\emph{Comput. Phys.
  Commun.} {\bf 253} (2020) 107198}
  [\href{https://arxiv.org/abs/1704.06650}{{\tt arXiv:1704.06650}}].

\bibitem{Dixon:1974ss}
J.A.~Dixon and J.C.~Taylor, \emph{{Renormalization of wilson operators in gauge
  theories}}, \href{http://dx.doi.org/10.1016/0550-3213(74)90598-7}{\emph{Nucl.
  Phys. B} {\bf 78} (1974) 552}.

\bibitem{Joglekar:1975nu}
S.D.~Joglekar and B.W.~Lee, \emph{{General Theory of Renormalization of Gauge
  Invariant Operators}},
  \href{http://dx.doi.org/10.1016/0003-4916(76)90225-6}{\emph{Annals Phys.}
  {\bf 97} (1976) 160}.

\bibitem{Joglekar:1976eb}
S.D.~Joglekar, \emph{{Local Operator Products in Gauge Theories. 1.}},
  \href{http://dx.doi.org/10.1016/0003-4916(77)90014-8}{\emph{Annals Phys.}
  {\bf 108} (1977) 233}.

\bibitem{Joglekar:1976pe}
S.D.~Joglekar, \emph{{Local Operator Products in Gauge Theories. 2.}},
  \href{http://dx.doi.org/10.1016/0003-4916(77)90170-1}{\emph{Annals Phys.}
  {\bf 109} (1977) 210}.

\bibitem{Kluberg-Stern:1974nmx}
H.~Kluberg-Stern and J.B.~Zuber, \emph{{Renormalization of Nonabelian Gauge
  Theories in a Background Field Gauge. 1. Green Functions}},
  \href{http://dx.doi.org/10.1103/PhysRevD.12.482}{\emph{Phys. Rev. D} {\bf 12}
  (1975) 482}.

\bibitem{Kluberg-Stern:1975ebk}
H.~Kluberg-Stern and J.B.~Zuber, \emph{{Renormalization of Nonabelian Gauge
  Theories in a Background Field Gauge. 2. Gauge Invariant Operators}},
  \href{http://dx.doi.org/10.1103/PhysRevD.12.3159}{\emph{Phys. Rev. D} {\bf
  12} (1975) 3159}.

\bibitem{Henneaux:1993jn}
M.~Henneaux, \emph{{Remarks on the renormalization of gauge invariant operators
  in Yang-Mills theory}},
  \href{http://dx.doi.org/10.1016/0370-2693(93)91187-R}{\emph{Phys. Lett. B}
  {\bf 313} (1993) 35} [\href{https://arxiv.org/abs/hep-th/9306101}{{\tt
  hep-th/9306101}}].

\bibitem{BluemleinLL}
J.~Bl\"umlein, P.~Marquard, K.~Sch\"onewald and C.~Schneider,
  \emph{{Proceedings of this conference}},  2022.

\bibitem{TongZhiLL}
T.Z.~Yang, T.~Gehrmann and A.~von~Manteuffel, \emph{{Proceedings of this
  conference}},  2022.

\bibitem{Falcioni:2022fdm}
G.~Falcioni and F.~Herzog, \emph{{Renormalization of gluonic leading-twist
  operators in covariant gauges}},
  \href{http://dx.doi.org/10.1007/JHEP05(2022)177}{\emph{JHEP} {\bf 05} (2022)
  177} [\href{https://arxiv.org/abs/2203.11181}{{\tt arXiv:2203.11181}}].

\bibitem{Nakanishi:1966zz}
N.~Nakanishi, \emph{{Covariant Quantization of the Electromagnetic Field in the
  Landau Gauge}}, \href{http://dx.doi.org/10.1143/PTP.35.1111}{\emph{Prog.
  Theor. Phys.} {\bf 35} (1966) 1111}.

\bibitem{Lautrup:1967zz}
B.~Lautrup, \emph{{Canonical Quantum Electrodynamics in covariant Gauges}}, .

\bibitem{Binosi:2013cea}
D.~Binosi and A.~Quadri, \emph{{Anti-BRST symmetry and background field
  method}}, \href{http://dx.doi.org/10.1103/PhysRevD.88.085036}{\emph{Phys.
  Rev. D} {\bf 88} (2013) 085036} [\href{https://arxiv.org/abs/1309.1021}{{\tt
  arXiv:1309.1021}}].

\bibitem{Becchi:1975nq}
C.~Becchi, A.~Rouet and R.~Stora, \emph{{Renormalization of Gauge Theories}},
  \href{http://dx.doi.org/10.1016/0003-4916(76)90156-1}{\emph{Annals Phys.}
  {\bf 98} (1976) 287}.

\bibitem{Tyutin:1975qk}
I.V.~Tyutin, \emph{{Gauge Invariance in Field Theory and Statistical Physics in
  Operator Formalism}},  \href{https://arxiv.org/abs/0812.0580}{{\tt
  arXiv:0812.0580}}.

\bibitem{Curci:1976bt}
G.~Curci and R.~Ferrari, \emph{{On a Class of Lagrangian Models for Massive and
  Massless Yang-Mills Fields}},
  \href{http://dx.doi.org/10.1007/BF02729999}{\emph{Nuovo Cim. A} {\bf 32}
  (1976) 151}.

\bibitem{Ojima:1980da}
I.~Ojima, \emph{{Another BRS Transformation}},
  \href{http://dx.doi.org/10.1143/PTP.64.625}{\emph{Prog. Theor. Phys.} {\bf
  64} (1980) 625}.

\bibitem{Baulieu:1981sb}
L.~Baulieu and J.~Thierry-Mieg, \emph{{The Principle of BRS Symmetry: An
  Alternative Approach to Yang-Mills Theories}},
  \href{http://dx.doi.org/10.1016/0550-3213(82)90454-0}{\emph{Nucl. Phys. B}
  {\bf 197} (1982) 477}.

\bibitem{Zinn-Justin:1974ggz}
J.~Zinn-Justin, \emph{{Renormalization of Gauge Theories}},
  \href{http://dx.doi.org/10.1007/3-540-07160-1_1}{\emph{Lect. Notes Phys.}
  {\bf 37} (1975) 1}.

\bibitem{DeWitt:1967ub}
B.S.~DeWitt, \emph{{Quantum Theory of Gravity. 2. The Manifestly Covariant
  Theory}}, \href{http://dx.doi.org/10.1103/PhysRev.162.1195}{\emph{Phys. Rev.}
  {\bf 162} (1967) 1195}.

\bibitem{Honerkamp:1971sh}
J.~Honerkamp, \emph{{Chiral multiloops}},
  \href{http://dx.doi.org/10.1016/0550-3213(72)90299-4}{\emph{Nucl. Phys. B}
  {\bf 36} (1972) 130}.

\bibitem{Honerkamp:1972fd}
J.~Honerkamp, \emph{{The Question of invariant renormalizability of the
  massless Yang-Mills theory in a manifest covariant approach}},
  \href{http://dx.doi.org/10.1016/0550-3213(72)90063-6}{\emph{Nucl. Phys. B}
  {\bf 48} (1972) 269}.

\bibitem{Kallosh:1974yh}
R.E.~Kallosh, \emph{{The Renormalization in Nonabelian Gauge Theories}},
  \href{http://dx.doi.org/10.1016/0550-3213(74)90284-3}{\emph{Nucl. Phys. B}
  {\bf 78} (1974) 293}.

\bibitem{Arefeva:1974jv}
I.Y.~Arefeva, L.D.~Faddeev and A.A.~Slavnov, \emph{{Generating Functional for
  the s Matrix in Gauge Theories}},
  \href{http://dx.doi.org/10.1007/BF01038094}{\emph{Teor. Mat. Fiz.} {\bf 21}
  (1974) 311}.

\bibitem{Sarkar:1974db}
S.~Sarkar, \emph{{Mixing of Operators in Wilson Expansions}},
  \href{http://dx.doi.org/10.1016/0550-3213(74)90428-3}{\emph{Nucl. Phys. B}
  {\bf 82} (1974) 447}.

\bibitem{Sarkar:1974ni}
S.~Sarkar and H.~Strubbe, \emph{{Anomalous Dimensions in Background Field
  Gauges}}, \href{http://dx.doi.org/10.1016/0550-3213(75)90633-1}{\emph{Nucl.
  Phys. B} {\bf 90} (1975) 45}.

\bibitem{tHooft:1975uxh}
G.~'t~Hooft, \emph{{The Background Field Method in Gauge Field Theories}},  in
  \emph{{12th Annual Winter School of Theoretical Physics}}, 1975.

\bibitem{Grisaru:1975ei}
M.T.~Grisaru, P.~van~Nieuwenhuizen and C.C.~Wu, \emph{{Background Field Method
  Versus Normal Field Theory in Explicit Examples: One Loop Divergences in S
  Matrix and Green's Functions for Yang-Mills and Gravitational Fields}},
  \href{http://dx.doi.org/10.1103/PhysRevD.12.3203}{\emph{Phys. Rev. D} {\bf
  12} (1975) 3203}.

\bibitem{Abbott:1980hw}
L.F.~Abbott, \emph{{The Background Field Method Beyond One Loop}},
  \href{http://dx.doi.org/10.1016/0550-3213(81)90371-0}{\emph{Nucl. Phys. B}
  {\bf 185} (1981) 189}.

\bibitem{Abbott:1981ke}
L.F.~Abbott, \emph{{Introduction to the Background Field Method}}, {\emph{Acta
  Phys. Polon. B} {\bf 13} (1982) 33}.

\bibitem{deVries:2019nsu}
J.~de~Vries, G.~Falcioni, F.~Herzog and B.~Ruijl, \emph{{Two- and three-loop
  anomalous dimensions of Weinberg\textquoteright{}s dimension-six CP-odd
  gluonic operator}},
  \href{http://dx.doi.org/10.1103/PhysRevD.102.016010}{\emph{Phys. Rev. D} {\bf
  102} (2020) 016010} [\href{https://arxiv.org/abs/1907.04923}{{\tt
  arXiv:1907.04923}}].

\bibitem{Moch:2021qrk}
S.~Moch, B.~Ruijl, T.~Ueda, J.A.M.~Vermaseren and A.~Vogt, \emph{{Low moments
  of the four-loop splitting functions in QCD}},
  \href{http://dx.doi.org/10.1016/j.physletb.2021.136853}{\emph{Phys. Lett. B}
  {\bf 825} (2022) 136853} [\href{https://arxiv.org/abs/2111.15561}{{\tt
  arXiv:2111.15561}}].

\bibitem{Freedman:1974gs}
D.Z.~Freedman, I.J.~Muzinich and E.J.~Weinberg, \emph{{On the Energy-Momentum
  Tensor in Gauge Field Theories}},
  \href{http://dx.doi.org/10.1016/0003-4916(74)90448-5}{\emph{Annals Phys.}
  {\bf 87} (1974) 95}.

\bibitem{Freedman:1974ze}
D.Z.~Freedman and E.J.~Weinberg, \emph{{The Energy-Momentum Tensor in Scalar
  and Gauge Field Theories}},
  \href{http://dx.doi.org/10.1016/0003-4916(74)90040-2}{\emph{Annals Phys.}
  {\bf 87} (1974) 354}.

\bibitem{Nogueira:1991ex}
P.~Nogueira, \emph{{Automatic Feynman graph generation}},
  \href{http://dx.doi.org/10.1006/jcph.1993.1074}{\emph{J. Comput. Phys.} {\bf
  105} (1993) 279}.

\bibitem{Misiak:1994zw}
M.~Misiak and M.~Munz, \emph{{Two loop mixing of dimension five flavor changing
  operators}},
  \href{http://dx.doi.org/10.1016/0370-2693(94)01553-O}{\emph{Phys. Lett. B}
  {\bf 344} (1995) 308} [\href{https://arxiv.org/abs/hep-ph/9409454}{{\tt
  hep-ph/9409454}}].

\end{thebibliography}\endgroup

\end{document}